# Multiscale assay of unlabeled neurite dynamics using phase imaging with computational specificity (PICS)

**Author List:** Mikhail E. Kandel[1,2,†], Eunjae Kim[1,2,†], Young Jae Lee[1,3], Gregory Tracy[5], Hee Jung Chung[3,5], Gabriel Popescu[1,2,4*]

1. Beckman Institute, University of Illinois at Urbana-Champaign, Urbana, IL 61801, USA
2. Department of Electrical and Computer Engineering, University of Illinois at Urbana-Champaign, Urbana, IL 61801, USA
3. Neuroscience Program, University of Illinois at Urbana-Champaign, Urbana, IL 61801, USA
4. Department of Bioengineering, University of Illinois at Urbana-Champaign, Urbana, IL 61801, USA
5. Department of Molecular Integrated Physiology, University of Illinois at Urbana Champaign, Urbana, IL 61801, USA

[†] Equal contributions

*Correspondence to:

Gabriel Popescu, 4055 Beckman Institute, 405 North Mathews Ave, Urbana, Illinois 61801, (217) 333-4840, gpopescu@illinois.edu

**Abstract:**

Primary neuronal cultures have been widely used to study neuronal morphology, neurophysiology, neurodegenerative processes, and molecular mechanism of synaptic plasticity underlying learning and memory. Yet, the unique behavioral properties of neurons make them challenging to study - with phenotypic differences expressed as subtle changes in neuronal arborization rather than easy to assay features such as cell count. The need to analyze morphology, growth, and intracellular transport has motivated the development of increasingly sophisticated microscopes and image analysis techniques. Due to its high-contrast, high-specificity output, many assays rely on confocal fluorescence microscopy, genetic methods, or antibody staining techniques. These approaches often limit the ability to measure quantitatively dynamic activity such as intracellular transport and growth. In this work, we describe a method for label-free live-cell cell imaging with antibody staining specificity by estimating the associated fluorescent signals via quantitative phase imaging and deep convolutional neural networks. This computationally inferred fluorescence image is then used to generate a semantic segmentation map, annotating subcellular compartments of live unlabeled neural cultures. These synthetic fluorescence maps were further applied to study the time-lapse development of hippocampal neurons, highlighting the relationships between the cellular dry mass production and the dynamic transport activity within the nucleus and neurites. Our implementation provides a high-throughput strategy to analyze neural network arborization dynamically, with high specificity and without the typical phototoxicity and photobleaching limitations associated with fluorescent markers.

**Introduction**

Neuronal branching and arborization provide a phenotypic marker for cellular viability and neurogenerative diseases[1-4]. While phase contrast microscopy can be used for studying neuronal cultures[5], the resulting images often struggle to differentiate between neurons and glia[6], and offer little beyond qualitative morphological information. Due to the need for chemical specificity, fluorescence-based techniques have become the main tools in neuroscience[7]. For example, confocal microscopy used in combination with immunostaining can reliably study axonal growth and dendritic branching[8]. When cells are fixed, dynamic information is painstakingly extracted by recording images from different subpopulations at different times. These challenges motivated the use of fluorescence protiens[9], which in turn introduces restrictions, such as phototoxicity[10], throughput[11], with the transfection process often hampering experiments with limited timeframes[12].

Quantitative phase imaging (QPI)[13-15], which derives morphology information from the scattered light by unlabeled specimens, offers a non-destructive method for studying cellular dynamics. This is accomplished by using interferometry to extract intrinsic information about the scattering potential associated with the object. As the scattering potential is invariant to the imaging system, it can be used to measure physical parameters such as the dry mass content of the cell[16-19]. In a broader context, QPI techniques promise to improve the image sensitivity to nanostructures[20], facilitate 3D imaging[21-23], and reduce observational bias due to staining[24,25] and fluorescent labels[26-30]. Although the quantitative phase measurement adds new infomation to the transmitted light signal, it nevertheless lacks molecular specificity. However, recent progress in artificial intelligence offers a potential solution. With the advent of artificial intelligence techniques based on deep convolutional neural networks[31], a new image-to-image translation[32,33] strategy has emerged where artificial stains can be inferred from the quantitative phase image itself[34]. Exploiting the high sensitivity to structure and quantitative information, phase imaging with computational specificity (PICS) was used to characterize the dry mass growth rate of subcellular compartments[35].

In this work, we show that PICS can be used to measure the arborization process in unlabeled neural cultures, over multiple days, nondestructively (Fig. 1). Our method consists of high-sensitivity QPI as well as end-to-end image analysis to infer the fluorescence intensity for Tau & MAP2[36], commonly used to identify axons and dendrites. These PICS derived fluorescent images are then used to label the neuronal compartments with subcellular specificity. In order to capture subcellular growth and intracellular transport, we apply the PICS-derived semantic segmentation maps to the dry mass density images rendered by QPI. We validated our assay by performing high-content screening of early-stage hippocampal cultures and observed several remarkable relationships between dry mass transport and growth in neurons.

**Results**

*<u>PICS Workflow for Semantic Segmentation</u>*

PICS combines high-sensitivity, temporally stable QPI with deep learning to estimate fluorescence stains from unlabeled specimens. (Fig. 1). The inferred fluorescence signal is processed to generate semantic segmentation maps which are then used to analyze the transport and growth of cellular dry mass. To acquire the quantitative phase images, we use gradient light interference microscopy (GLIM), which measures the optical path length shifts associated with the specimen in a differential interference contrast (DIC) geometry.

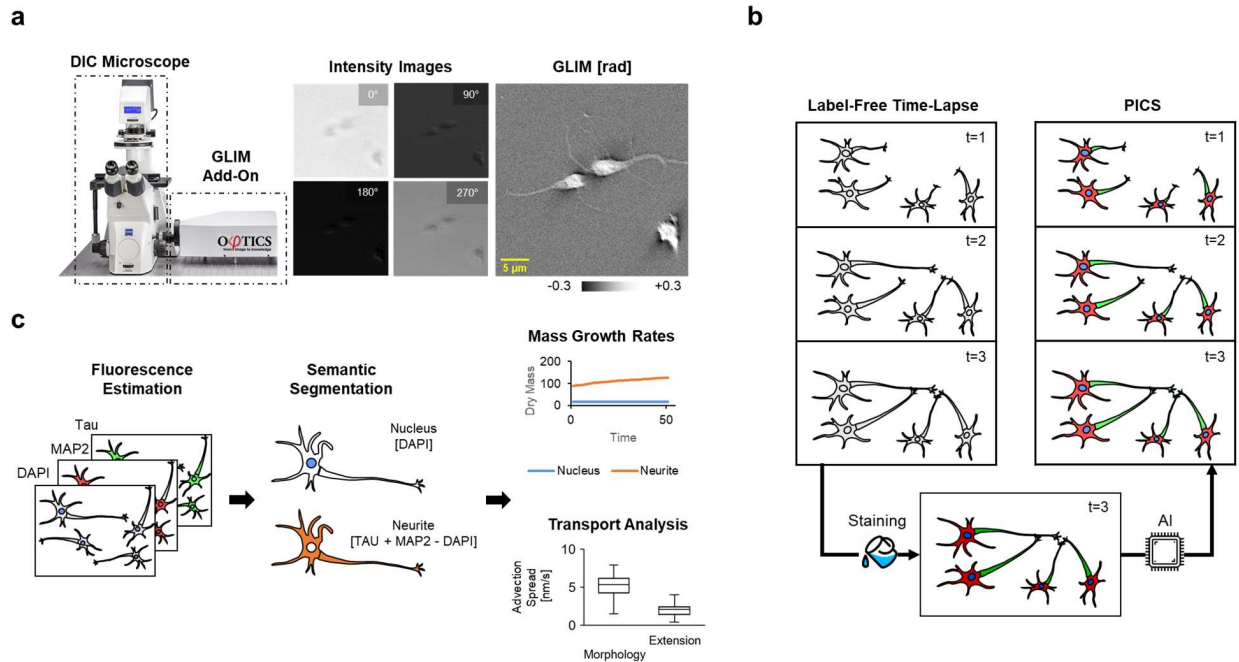

**Figure 1. Phase Imaging with Computational Specificity (PICS) for measuring growth and transport during neural arborization. a,** The GLIM system upgrades a conventional differential interference contrast microscope with quantitative phase imaging capabilities **b,** Hippocampal cultures were imaged over 41 hours for time-lapse analysis (20x/0.8). After the recording, neurons were fixed and stained with antibodies for Tau and MAP2 to obtain co-localized phase and fluorescence images. An additional nucleus (DAPI-like) channel is provided by manual annotation. To recover time-lapse data with specificity to antibodies, deep convolutional neural networks trained on the fixed cells were used to infer the fluorescent signals on live cells. **c,** PICS (inferred fluorescence) maps for Tau, MAP2, and nuclei created a three-channel semantic segmentation map, labeling the image as "background", "nucleus" and "neurite". The segmentation map is then used to characterize the neural growth rates and intracellular mass transport.

Following the procedure in Fig. 2 (see Supplementary Note 1 for details), we acquire four intensity frames corresponding to $\pi/2$ offsets between the two laterally shifted beams in DIC. The interference between these two beams reveals the derivative of the phase map along the direction of the shift. The QPI map is obtained by integrating this derivative using a Hilbert transform (Supplementary Fig. 1c). To identify axons and dendrites, we performed antibody staining for the Tau and MAP2 proteins, respectively[37]. As the detection light paths in transmission and epi-fluorescence are shared, it is straightforward to acquire co-localized fluorescence images using the same camera detector. Furthermore, we included a PICS DAPI label by manually annotating the nuclei in the phase images.

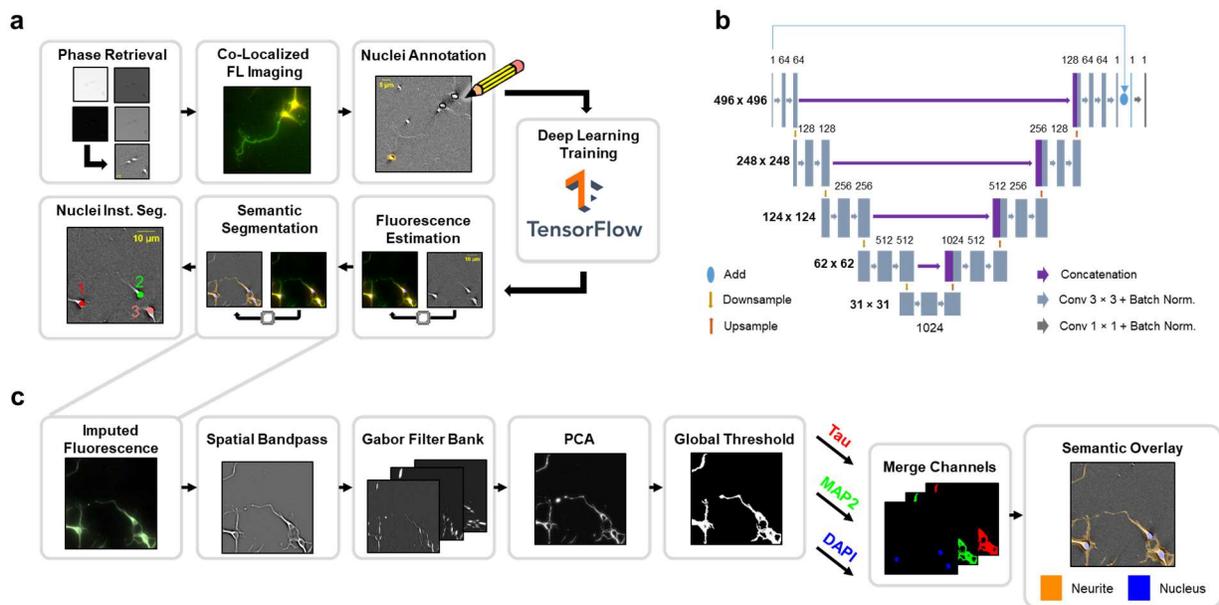

**Figure 2. Workflow for automatically annotating GLIM images. a,** Co-localized GLIM and FL images were digitally registered. The phase data were reconstructed using a Hilbert transform integration technique. Nuclei were manually annotated in ImageJ to simulate DAPI staining. A deep convolutional neural network was trained using measured data to reproduce the fluorescence channels (Tau/MAP2/DAPI) from the label-free GLIM image. To analyze dry mass growth rates, we performed inference on the unstained time-lapse sequence. The estimated fluorescence is converted into a semantic segmentation map with labels for neurite, nucleus, and background. Finally, connected component analysis on the nucleus associated regions is used to produce instance segmentation to count the number of nuclei in each field of view. **b,** Images are trained using a U-Net architecture consisting of 64 filters at the first layer, and a Pearson correlation between the actual and estimated fluorescence is used as the loss function. **c,** Estimated

fluorescence maps are processed to obtain a binary map by applying a spatial bandpass, generating textures using a Gabor filter bank, and reducing the dimensionality with principal component analysis. Finally, a threshold is applied to binarize the image for each channel. These three channels are then merged to form an annotated image with labels for the neurite, nucleus, and background.

These co-localized fluorescence images are used to train neural networks that estimate the fluorescence image from the transmitted light GLIM image. We note that while the four frames that constitute the GLIM image took approximately 200 ms to acquire, each fluorescent channel was acquired by averaging a total of ten images at 700 ms exposure each, with 2x2 binning. Thus, we found that fluorescence microscopy was 70x slower than phase imaging. While antibody staining is often faint and depends on protein expression, the signal in transmitted light imaging can be modulated by simply increasing the strength of the illumination (with no risk of photobleaching). These results highlight an important throughput advantage of synthetic rather than physical staining.

Next, we use the neural networks trained on fixed, antibody stained cells to perform inference on the unstained live neurons. As shown in Fig. 2c, the estimated fluorescence signal is converted into a semantic segmentation map through a series of image processing steps. For each estimated fluorescence image, we perform a spatial bandpass[20] to remove low frequencies and generate a series of variants of the image using a Gabor filter bank[38]. The resulting set of images highlights the textural information and contains many values at each pixel. The parameters are reduced to a single channel by principal component analysis (PCA)[39], with a global threshold applied to binarize the image. Compared to simpler global thresholding approaches that rely on histogram analysis, Gabor filters capture textural information which allows for more accurate segmentation.

This procedure is repeated to generate a binary map for each channel (Tau, MAP2). To merge the channels into a three-category semantic segmentation map ("nucleus", "neurite", "background") we take the nuclear binary map, add the Tau & MAP2 binary images, and assign the "background" label to the rest of the pixels. We merged the Tau and MAP2 channels to increase signal-to-noise and study the growth and dynamics of both axons and dendrites. Finally, we perform instance segmentation to count the nuclei using simple connected-component analysis (CCA)[40] on the binary nucleus-associated labels. We validated our cell counting technique by comparing manual to automatic nuclear counts on the first time point. We obtained a Pearson correlation coefficient of 0.94 between the two techniques, with the principle disagreement stemming from cell clusters and glia cells. The manual cell count took roughly three hours to perform, while the CCA-based method completed in under a minute.

*Neural Network Architecture*

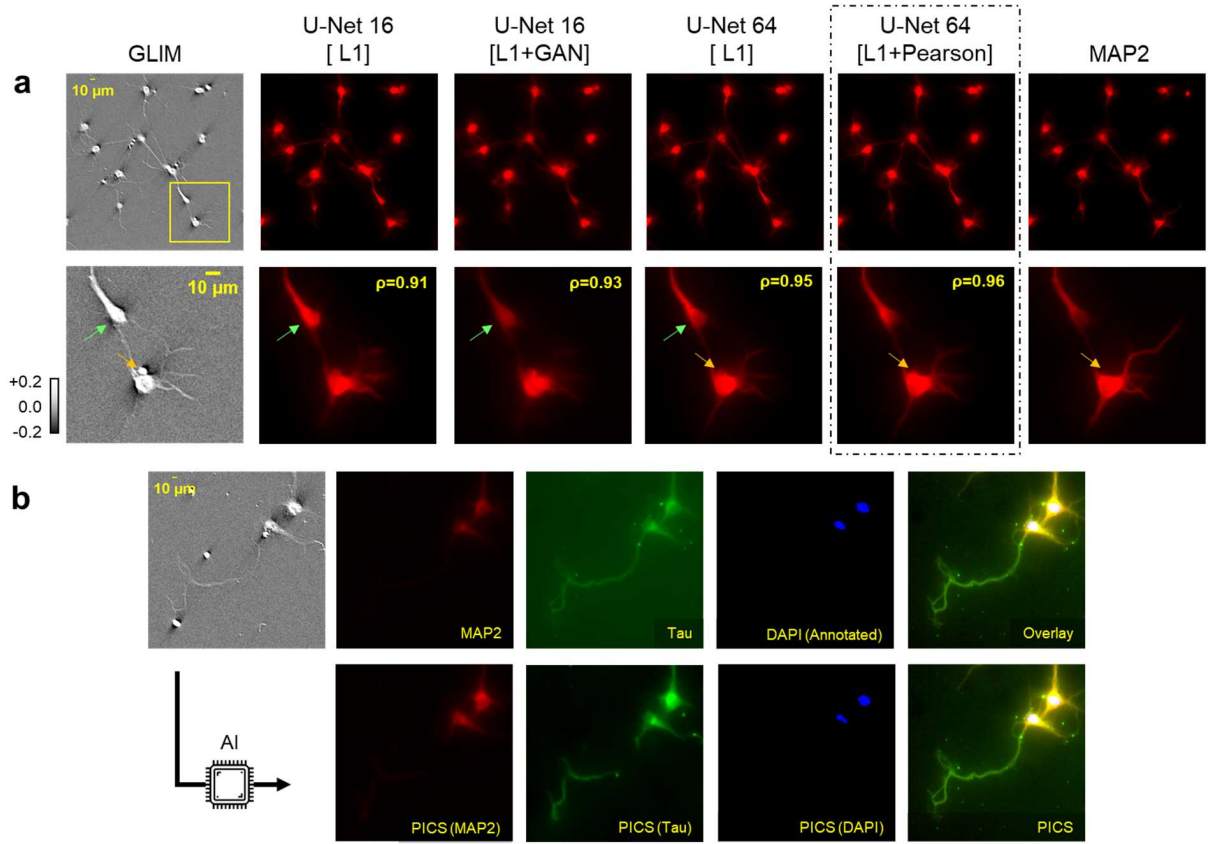

**Figure 3. Deep convolutional neural network training. a,** To investigate the effect of neural network architecture loss functions on image quality, we measured the Pearson correlation between the actual and estimate fluorescence images. As a baseline for performance, we took a U-Net constructed with a reduced parameter set consisting of 16 initial filter elements trained against an L1 loss function. Compared with the conventional U-Net training scheme, we found that a GAN improved performance by suppressing the overestimation of MAP2 concentration in the nucleus (green arrows). Further improvements came from expanding the number of filters in the U-Net architecture, with the GAN removed for faster training time (U-Net 64). Switching the loss function to the Pearson correlation further improved performance with the resulting images suppressing artifacts, such as unwanted debris around the nuclei (U-Net 64, Pearson, yellow arrow). **b,** Actual vs estimated fluorescence images showed strong agreement, with only a slight mismatch in long axons (20x/0.8).

Central to our approach is the estimation of fluorescence images from the unlabeled phase maps[35]. To accomplish this we performed image-to-image translation using a deep convolutional neural network[32]. Deep convolutional neural networks are well suited to this task as they combine the texture-based techniques with added nonlinearities hidden in their deep layers. We used the U-Net architecture, which includes downsampling, and upsampling paths to efficiently integrate both textural and contextual information[41].

To arrive at the final architecture used to estimate our fluorescent tags, we tuned the number of filter banks and trained the network with an unconventional loss function. The results of these experiments are summarized in Fig. 3. We selected the architecture in[35] as it is known to have near real-time inference performance on commodity computing hardware, with a filter bank of sixteen filter elements at the input layer (Fig. 3, "U-Net 16"). All networks were trained against an L1 reconstruction loss function[42]. While able to reproduce much of the morphology of the MAP2 signal ($\rho=0.91$), the network was poor at capturing variations of protein within cells, often being unable to reject the MAP2 signal within the nucleus (Fig. 3, green arrows). To bias the training procedure but maintain the same inference time, we introduced a generative adversarial network (GAN)[43] training scheme where the U-Net is taken as the generator, and PatchGan was used as the discriminator[32]. Although performance improved as evidenced by a reduction in the overstaining of the top neuron and an improved correlation coefficient (Fig. 3, green arrow, $\rho=0.93$), the nucleus within the top neuron appeared to be distorted. Attributing these defects to the tendency of GANs to introduce features where none exist[44], we instead removed the GAN and expanded the initial filter bank size to 64 elements (U-Net 64). With more filters, this scheme is more successful in capturing subtle details, and the nuclear vs non-nuclear area is clearly delimited in the top neuron (Fig. 3, green arrow). Finally, we modified the loss function to use the Pearson correlation between the actual and estimated fluorescence image, which is our ultimate quality metric[45]. The resulting network was able to delineate cellular morphology and discriminate between cell bodies and cell shaped debris (Fig. 3, orange arrow). Qualitatively, the estimated fluorescence signal shows a strong resemblance to the actual fluorescence signal (Fig. 3b). Unlike the real stain, the estimated fluorescent signal avoids autofluorescence and other unwanted sources of noise, especially in the background (Fig. 3b, Tau).

## *Time-Lapse Antibody Staining Prediction*

Among the chief advantages of the proposed method is the ability to perform live-cell imaging with the specificity of computationally inferred stains that would otherwise require fixation. This is especially true of antibody-based staining techniques, which require the cell to be cross-linked, and made permeable, a procedure that is incompatible with live-cell imaging[36]. To overcome this challenge, we performed time-lapse GLIM imaging over a period of a few days and stained the cells at the end of the experiment for neural network training. The characterization of primary hippocampal neuronal arborization has been recognized as valuable for applications including drug discovery and toxicity screening[46,47] (Fig. 4).

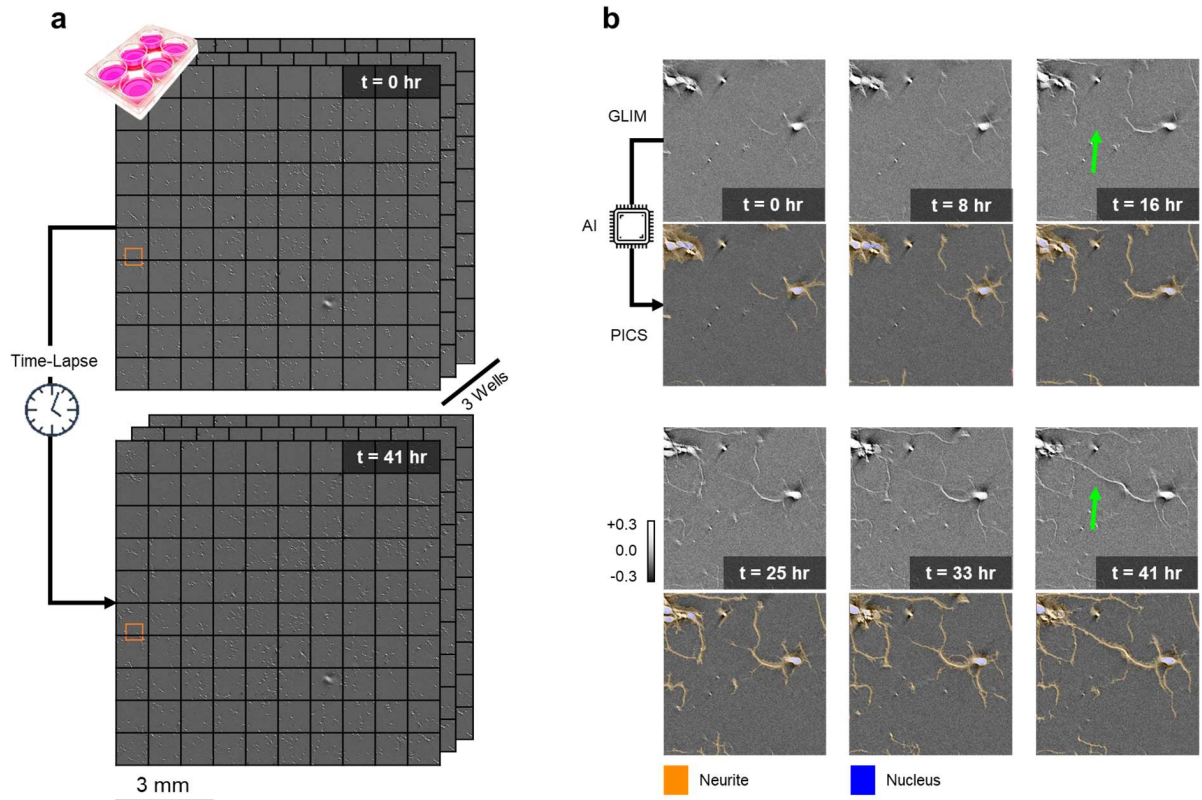

**Figure 4. PICS provides antibody specificity to unlabeled live cells. a,** 41 hours of time-lapse mosaic imaging was performed on a six-well plate, with three mosaics scans (10 x 10 images, 5 x 5 mm$^2$) acquired at each time point. The plate was digitized in under two minutes. **b,** Zoomed-in portion of a representative region (orange box) shows increasing arborization as the neurons connect (20x/0.8).

We imaged early-stage hippocampal neurons consisting of three 5 x 5 mm$^2$ regions under a 20x/0.8 objective in a multiwell plate that is typically used to image several preparations in parallel (Fig. 4). Each well was imaged by GLIM in roughly half a minute (see Supplementary Fig. 2 for details). As per our protocol outlined in Fig. 1b, at the end of the experiment, the cells were fixed, immunostained, and imaged by both GLIM and fluorescence microscopy. Next, we trained the neural networks to perform a remapping from the GLIM image to the co-localized fluorescence images, which are then processed into a semantic segmentation map (Fig. 2,

Supplementary Fig. 3). Importantly, with an efficient GLIM light budget, we can trade exposure time for reduced light intensity and reduced toxicity[10], which is especially important for imaging highly sensitive cells such, as neurons[48].

We observed that dendrites show steadier growth when compared to axons, which appear to actively search for connections. We also witnessed that, when a cluster is approached by an axon, it will rapidly divert its dendrite to form a connection[49] (Fig. 4, hour 33 to hour 41, green arrow). These behaviors, as well as others, are shown in Supplementary Movies 1,2,3.

## *Mass Transport Analysis*

Cells often express phenotypes that are related to changes in the transport of cellular cargo. We stress that this is a dynamic activity requiring live-cell imaging and time-lapse observation. To analyze the transport of cellular mass we employed dispersion-relation phase spectroscopy (DPS), which reports on the transport rate of dry mass at different spatial scales[50].

In DPS, cellular material is understood to be governed by the diffusion-advection equation, so that it is possible to measure the spread in advection velocities associated with active cellular transport. DPS has been previously applied to measuring transport in live cells[51,52]. However, here we used DPS in combination with PICS segmentation, which allows for studying dynamic transport in subcellular compartments. Using the high-throughput AI-based segmentation, we measured the transport associated with 300 fields of view or 4,679 neurons. Our analysis method is discussed in Methods with a pictorial representation of the processing steps shown in Supplementary Fig. 4.

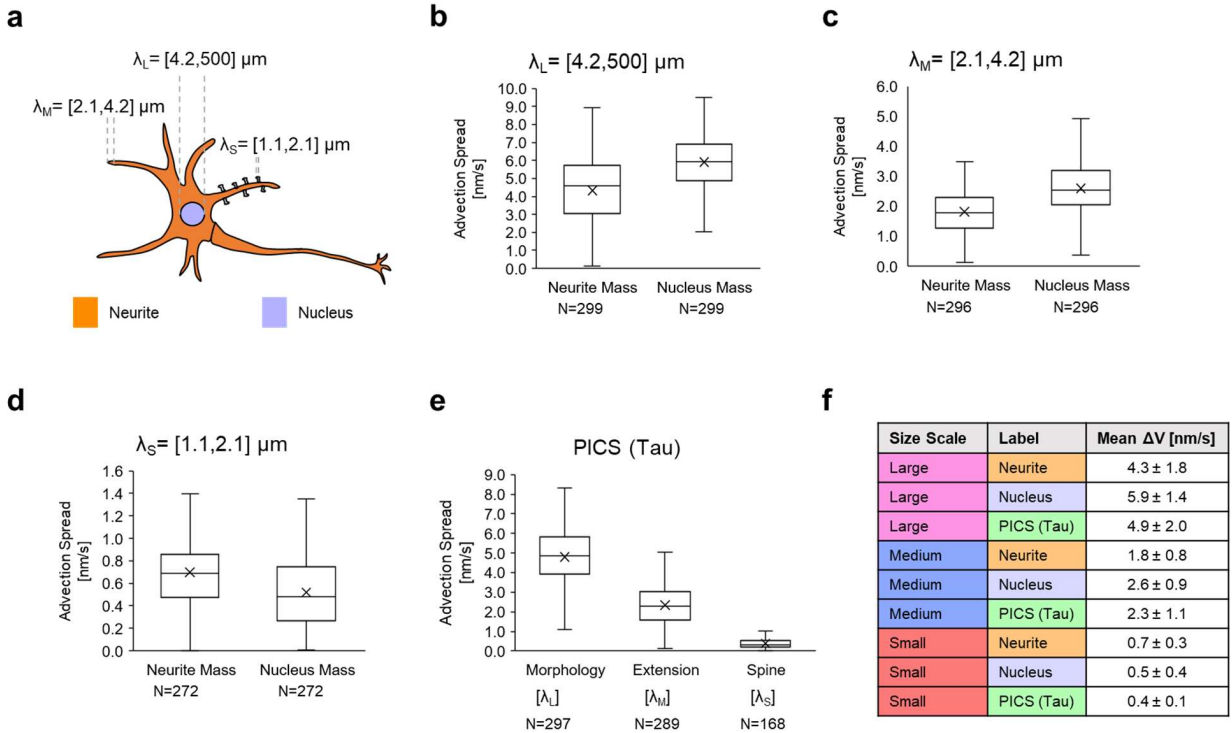

**Figure 5. Transport analysis shows substantial nuclear reconfiguration during neuronal arborization. a,** We applied DPS at three spatial scales, corresponding to dendritic spines ($\lambda_{Small}$=[1.1,2.1] μm), neurites ($\lambda_{Medium}$=[2.1,4.2] μm), and cellular morphology ($\lambda_{Large}$=[4.2,500] μm). **b-d** By using the semantic segmentation maps generated from PICS, we collect the advection velocity spread corresponding to transport within the neurite and nucleus. Boxplots consist of all fields of view that were found to exhibit active transport, and all differences between categories were statically significant (Mann–Whitney U, p-value <0.05). **e,** Transport analysis was run directly on the inferred fluorescence channel. **f,** Summary of the dry mass advection spread distributions.

As shown in Fig. 5, we measured cellular dynamics at three spatial scales; with small scales ($\lambda_S$=[1.1,2.1] μm) corresponding to the size of dendritic spines and nucleoli, medium scales ($\lambda_M$=[2.1,4.2] μm) approximately at the width of the extensions, and large scales ($\lambda_L$=[4.2,500]) on the order of the cell size. Following the DPS procedure outlined in methods, we estimated the variance of the temporal power spectrum at each Fourier frequency in the image sequence. This procedure estimates the activity at each spatial mode (Fourier frequency), with a linear curve fit taken at three relevant spatial frequencies (Supplementary Fig. 4). We performed this analysis for the mass that falls within the neurite and nucleus by setting to zero the dry mass values that lie outside the categories. In this work, we looked at transport at relatively long temporal scales, meaning that fast-moving diffusive particles are unlikely to be captured between imaging intervals. Thus, we found that most fields of view were dominated by advection, *i.e.*, temporal bandwidth, $\Gamma$, linear in the spatial mode $q$ ($\Gamma \propto q$), and the few sequences that did not meet this model were excluded from our analysis.

Our results for dry-mass transport analysis are summarized in Table 1. As a plate read was performed every 48 minutes, we primarily capture the slow movement consistent with anterograde transport associated with cytoskeletal proteins[53]. We observed that at "large" size scale ($\lambda_L$) associated with cellular morphology, the nucleus associated mass transport exhibits a 37% higher spread in advective mass transport. Although the spread in mass transport decreases by roughly a factor of two in absolute terms, the nucleus associated mass has a 44% larger spread at scales corresponding to the width of the neuronal extensions ("medium" $\lambda_M$). These results hint that despite making up a small fraction of the total mass of the neuronal arbor (roughly 1/4), at scales comparable to cellular morphology and extensions, the nucleus exhibits a remarkable diversity of mass transport activity. This result is in contrast to one for neurites, which are relatively steady in their growth. Yet, at smaller scales ($\lambda_S$), this relationship becomes reversed with nucleus associated mass transport showing a 29% smaller spread in velocity coefficients. This is not surprising, as smaller scales include neuronal cargo and rapidly moving dendritic spines. All differences were found to be statically significant (Mann-Whitney U, p-value <0.05).

Next, we propose to use PICS to measure the transport of the antibody-associated protein rather than the cellular dry-mass. The method that extends the transport analysis to fluorescence data is referred to as dispersion-relation fluorescence spectroscopy[54]. However, with PICS, we can study the transport of specific molecular structures without labeling. We selected Tau, as abnormalities in its localization are associated with many cellular pathologies[55]. To calculate these parameters, we substituted the estimated fluorescence signal for the dry-mass and computed the DPS advection coefficients associated with the whole field of view. We stress that this type of observation is performed here for the first time, as antibody staining requires destructively fixing the cell. The distribution of transport coefficients is shown in Fig. 5e. We note that the DPS transport activity for the Tau protein inferred by PICS is somewhere between the neurite and nucleus activity for size scales, corresponding to cellular morphology ($\lambda_L$) and extensions ($\lambda_M$).

Further, the estimated Tau transport is comparably lower for size scales corresponding to dendritic spines ($\lambda_S$), which are known to contain less Tau protein when healthy[56].

## *Dry Mass Analysis*

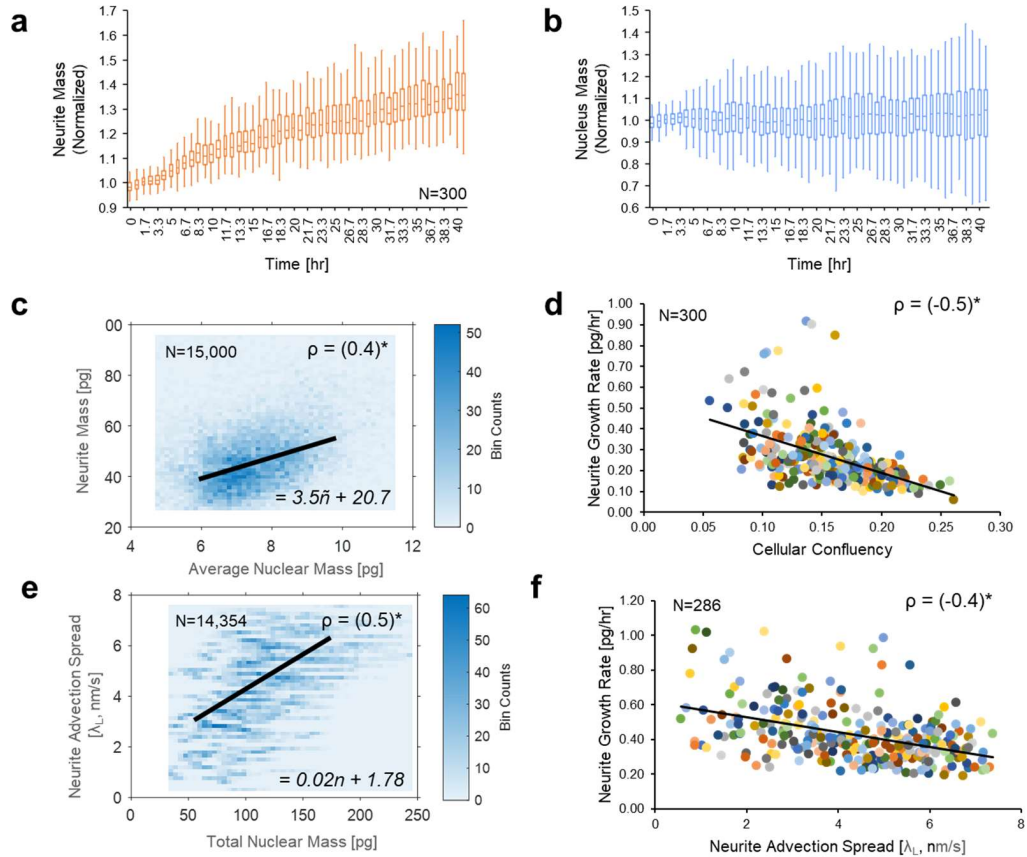

**Figure 6. Dry mass analysis for subcellular compartments reveal relationships between growth rates and transport. a-b,** The semantic segmentation maps track the dry mass of nuclei and neurites for 41 hours over 300 fields of view, corresponding to approximately four thousand neurons. Dry mass was segregated into nucleus and neurite components by summing the phase values in the GILM image within the label specified by the PICS map. To extract the relative mass change, these values were normalized by the average dry mass during the first five hours. While neurite dry mass shows steady growth, the behavior of the nucleus is more erratic, with no significant growth overall. **c,** Neurite mass appears to depend on average nuclear mass ($\rho=0.4$), with a nuclear count obtained from binary morphological operations at the start of the experiment. **d,** The neurite growth rate is negatively related to the cellular confluency ($\rho=-0.5$). Here we

compute confluency on a per-tile basis by looking at the fraction of the tile occupied by cellular material. **e,** Neurite transport behavior for $\lambda_L$ depends on the total nuclear mass ($\rho$=0.5). **f,** Neurite growth rate appears to be anti-correlated with advection spread ($\rho$=-0.4).

The PICS approach differentiates itself from previous efforts on synthetic staining[33] by taking a step further and analyzing the cellular dry mass – a measurement that is specific to quantitative phase imaging[14]. We used the Tau/MAP2 semantic segmentation to distinguish the dry mass associated with the nucleus from that of the neurites.

In Fig. 6a-b, we measure the normalized rate of change of the dry mass within each of the 300 mosaic tiles (Fig. 4a). The results indicate that the neurite mass shows a steady increase, while the growth associated with the nucleus is very low on average, with a high variance in time. To investigate this variability, we looked at the relationship between the average nuclear mass and neurite mass. We calculated the ratio between the per-tile nuclear-associated mass and the nuclear count to obtain the average nucleus mass within the field of view. We noticed an interesting linear relationship between the average nuclear mass and the average neurite mass (Fig. 6c), with the average neurite mass being roughly four times the mass of the nucleus ($\rho$ = 0.4). Additionally, we observed the expected relationship between neurite growth rates and cellular confluence, which shows that neurites exhibit growth inhibition as their density increases (Fig. 6d). For this calculation, the cellular confluence was measured on a per-tile basis at the start of the experiment, and the growth rate was determined by a linear fit.

Finally, we combined the semantic segmentation maps generated by PICS with advection coefficients and growth rates to highlight the interplay between nuclear mass and transport. For each tile, we obtained a coefficient for large scale transport ($\lambda_L$), following the procedure in the previous section. In total, we correlated 15,000 points from all the acquired data. In Fig. 6e, we observe a direct relationship between nuclear mass and transport in neurites ($\rho = 0.4$). In contrast, rates of nuclear mass change (growth) were found to be anti-correlated with transport ($\rho = -0.4$). These findings suggest that a more massive nucleus promotes fast traffic in the neurite. At the same time, the transport is slowed down when the neurites grow.

**Methods**

*<u>Dry Mass Calculation from Phase Images</u>*

QPI is governed by a scattering potential that describes how the incident wavefront is distorted by the object. Through a series of pioneering experiments measuring different solutions of proteins and lipids, it was found that dry-mass concentration and scattering potential are linearly related[57]. As the scattering potential occurs due to the refractive index difference with respect to the surrounding medium, QPI yields the non-aqueous ("dry") content of the cell.

In conventional microscopes such as DIC or phase-contrast, the relationship between the object's scattering potential and the recorded image is not linear[27]. In this case, the recorded image depends on the illumination as well as light propagating from out of focus planes[20,27]. QPI yields phase maps that are not corrupted by the two amplitudes of the interfering fields. Thus, the phase image within a given optical section provides access to the dry mass density, $\sigma$,

$$\sigma(\mathbf{r}) = \frac{\lambda}{2\pi\alpha} \phi(\mathbf{r}) \tag{1.1}$$

where $\lambda$ is the wavelength of light and $\alpha$ is the refractive index increment, and $\phi(\mathbf{r})$ is the quantitative phase image after integration, $\mathbf{r} = (x, y)$.

For correlative analysis, we estimate the rate of dry mass growth by using Savitsky-Golay based filtering[58] with the time points laying at the peripheries of the sequence omitted from analysis.

*Neural Network Training*

Table 1

|  | Training | Validation | Test | Total | Augmentation |
|---|---|---|---|---|---|
| Tau/MAP2 | 784 | 88 | 100 | 972 | x4 |
| DAPI | 640 | 80 | 80 | 800 | x2 |

We trained three separate networks to predict Tau, MAP2, and DAPI stains. We used U-Net for the main architecture as discussed in Results. The networks for Tau and MAP2 are trained with the co-localized phase and fluorescent image pairs. The outputs of these networks are predicted fluorescent images and are processed into the final semantic segmentation maps as outlined in Methods. The model losses for these networks were calculated as the weighted sum of mean absolute error and the Pearson correlation between the predicted fluorescent intensities and the ground truth fluorescent intensities[45]. Following a standard procedure, pixel values in the image pairs are normalized between 0 and 1, and images are translated by several pixels to account for shifts due to slight misalignments between fluorescent filter cubes[35].

The training, validation, and test split is summarized in Table 1. Initial data was acquired at 20x/0.8 corresponding to ~600 x 600 μm² size fields of view. One of the challenges with neural networks is training time, and here we ameliorated these concerns by downsampling the images by a factor of two using nearest neighbor resampling. This procedure is also performed to match the GLIM image to the fluorescence image which was acquired by 2x2 binning. Next tiles were divided into quarters. Thus, for estimating the Tau and MAP2 signal, we used 972 phase-fluorescent pairs each corresponding to an area of ~315 x 315 μm², with additional augmentation (rotation and flipping) performed on the fly during training.

The network for the DAPI-like nuclear annotation was trained to produce binary labels rather than estimated fluorescent intensities, and training was performed on a slightly different quantity of images (Table 1).

A commercial workstation running Gentoo[59] was equipped with two NVIDIA RTX 2070 GPUs and the networks were implemented using Keras[60] built on TensorFlow[61]. We used the ADAM optimizer[62] with a batch size of one on each GPU, effectively allowing the network to use larger training sizes within a single GPU. Training took 100 epochs and finished in 14 hours.

*Dispersion-Relation Phase Spectroscopy for Mass Transport Analysis*

Cellular mass transport can be analyzed by tracking moving objects or by analyzing spectral fluctuations in the image. We opt for the latter as the neuronal arbor forms a dense network that approaches a continuous dynamic system. In this work, we propose to use dispersion-relation phase spectroscopy (DPS)[50], which interprets the mass transport under the diffusion-advection equation. This method of analysis transforms a time-lapse image sequence into an estimate of the temporal bandwidth at each spatial mode, which can be used to calculate the diffusion/advection coefficients. The steps for this procedure are shown in Supplementary Fig. 4.

Following the derivation in[52], we assume that the inhomogeneous mass density (dry mass or Tau protein) within a semantic label, $\rho(\mathbf{q},t)$, follows the diffusion-advection equation,

$$\left(-D\mathbf{q}^2 + i\mathbf{q}\cdot\mathbf{v} - \frac{d}{dt}\right)\rho(\mathbf{q},t) = 0 \qquad (1.2)$$

Where $D$ is the diffusion coefficient, and $\mathbf{v}$ is directed velocity of the particles. We can write the velocity-averaged density as [52]

$$\langle\rho(\mathbf{q},\tau)\rangle_v = \exp(i\mathbf{q}\cdot\mathbf{v}_o\tau)\exp\left(-\left[D\mathbf{q}^2 + \Delta\mathbf{v}\cdot\mathbf{q}\right]\tau\right) \qquad (1.3)$$

where $\langle\ \rangle_v$ indicates ensemble averaging over the velocity distribution, $\mathbf{v}_o$ is the mean velocity, and $\Delta\mathbf{v}$ represents the variance of the distribution of velocities. The mean velocity, $\mathbf{v}_o$, is typically negligible, as mass transport is equally probably in both directions along a given line. Accordingly, the decay time at each spatial frequency, $\Gamma(\mathbf{q})$, characterizes the behavior of the sample, as

$$\Gamma(\mathbf{q}) = D\mathbf{q}^2 + \Delta\mathbf{v}\cdot\mathbf{q} \qquad (1.4)$$

In order to compute $\Gamma(\mathbf{q})$ efficiently, we evaluate it as the standard deviation of the temporal power spectrum. We note that while, in general, the second moment of a Lorentzian (Fourier transform of an exponential) is not defined, for these experiments we use finite imaging intervals which makes our integrals converge. For the zero-mean power spectra of interest here, the variance is computed as,

$$\Gamma^2(\mathbf{q}) = \frac{\int \omega^2 |\rho(\mathbf{q},\omega)|^2 \, d\omega}{\int |\rho(\mathbf{q},\omega)|^2 \, d\omega} = \frac{\int \left|\frac{d\rho(\mathbf{q},t)}{dt}\right|^2 dt}{\int |\rho(\mathbf{q},t)|^2 \, dt} \tag{1.5}$$

where $\rho(\mathbf{q},\omega)$ is the Fourier transform of $\rho(\mathbf{q},t)$ in time. Note that the differentiation property of the Fourier transform allows us to compute $\Gamma(\mathbf{q})$ directly in the time domain very quickly[63]. By curve fitting using the measured values of $\Gamma(\mathbf{q})$, it is possible to estimate $D$ and $\Delta\mathbf{v}$ for particular spatial frequency ranges. The isotropic intracellular transport is obtained by performing a radial average over $\Gamma(\mathbf{q})$. In this work, we develop a high-quality procedure for radial average that involves first mapping each point in $\Gamma$ to its radial coordinate $\|\mathbf{q}\|$ followed by an interpolation step (See Supplementary Fig. 4, Step 4: Radial Averaging).

As imaging was performed in 50-minute increments, we noticed that diffusive motion was negligible, *i.e.*, the quadratic $\mathbf{q}^2$ term was small. This makes the curve-fitting a straightforward linear fit. These operations were performed in an automated fashion using a script written in MATLAB.

For a more intuitive representation of the data, intervals of $q$ are converted into direct-space scales using the following relationship,

$$\lambda = \frac{2\pi}{q} \tag{1.6}$$

**Summary and Discussion**

By combining advanced microscopy with artificial intelligence, PICS allowed us to observe several competing trends during neuronal arborization. We found that for the hippocampal neurons studied in this work, there is a correlation between the average mass of a nucleus and neurites in the same neuron at a preferred dry mass ratio between the two (Fig. 6 c, roughly factor of 4). While it was hypothesized that neurons have intrinsic growth potentials in terms of cellular dry mass[64], this observation provides direct evidence that, during early network development, the mass within the nucleus balances the mass within the neurite. This hypothesis is further supported by the interplay of intracellular cellular transport and neurite growth. By limiting our imaging to slow time scales, we focused primarily on anteretrograde transport motion[65]. In Fig. 6 e, we observe that the dry mass associated with the nuclear mass is directly related to transport activity, with more nuclear mass leading to more active transport in the axons and dendrites. This is not surprising as more massive nuclei are expected to produce more cellular material. Our results highlight that mass transport and growth, two potentially independent phenomena, are anti-correlated in the neurites - where high growth rates appear accompanied by slower transport. This observation is in-line with the understanding that there exists a metabolic trade-off between cellular motion and vegetative growth[65-68]

The complexity of imaging neuronal clusters is primarily due to the photochemical sensitivity of the cells and the resolving power needed to detect fine structures such as dendrites within a developing arbor. To meet these challenges, our approach relies on machine learning, quantitative phase imaging, and transport analysis with few conventional analogs. At the core of our analysis scheme is the use of inferred fluorescence images which introduce specificity for subcellular compartments through a digital staining procedure (PICS)[35]. We found that inferring antibody stains requires much more computational complexity compared to simpler markers, such as DAPI. This motivated us to expand the deep-convolutional neuronal network and modify the loss function. Further, we found that the intrinsic variation of protein concentration associated with the MAP2/Tau antibodies made it difficult to generate a semantic segmentation map from the estimated fluorescence signal. This challenge was addressed by using texture-based thresholding to include local image features. In contrast to conventional tracking-based approaches for analyzing neuronal extensions (such as Sholl analysis[69]), to improve imaging throughput we used a continuous model to describing mass transport, which is invariant to annotation errors.

With a plate read performed within a few minutes, PICS enabled us to use label-free microscopes, which are much faster than their fluorescent counterparts. We found that it was roughly 70 times faster to computationally estimate, rather than acquire the fluorescent signal. Using the quantitative dry mass output from QPI and the specificity from AI, we were able to compute correlations between neuronal growth rates and intracellular transport, at unprecedented throughput, across thousands of cells.

While in this work we primarily investigated the relationship between nuclei and neurites, we expect that the extension of our mass growth & transport analysis to other stains such as those associated with the specific species of cargo or other cellular systems will be relatively straightforward. Ultimately, by using an interferometric module attached to a conventional DIC microscope and non-confocal epi-fluorescence imaging, we hope that PICS will be broadly adapted as an upgrade to existing microscopes.

**Data availability statement:**

The data that support the findings of this study are available upon reasonable request.

**Availability of computer code and algorithms:**

The code and computer algorithms that support the findings of this study are available from the corresponding author upon reasonable request.

**Acknowledgments:**

This work is supported by NSF 0939511 (G.P.), R01GM129709 (G.P.), R01 CA238191 (G.P.), R43GM133280-01 (G.P.), R01NS083402 (H.J.C), and R01NS097610 (H.J.C). M.E.K. is supported by a fellowship from the Miniature Brain Machinery Program at UIUC (NSF, NRT-UtB, 1735252).

**Contributions:**

M.E.K. designed and performed imaging experiments. G.T dissected cortices from rat E18 embryos. Y.J.L. cultured and stained hippocampal neurons. H.J.C. contributed reagents and animals. E.K. developed and trained the neural network. M.E.K. and E.K. performed analysis. M.E.K., E.K., and G.P. wrote the manuscript. G.P. supervised the project.

**Conflict of Interest:**

G.P. has a financial interest in Phi Optics, Inc., a company developing GLIM and PICS for materials and life science applications. The remaining authors declare no competing interests.

Supplementary information for

**Mutiscale assay of unlabeled neurite dynamics using phase imaging with computational specificity (PICS)**

# Supplementary Note 1: Gradient light interference microscopy

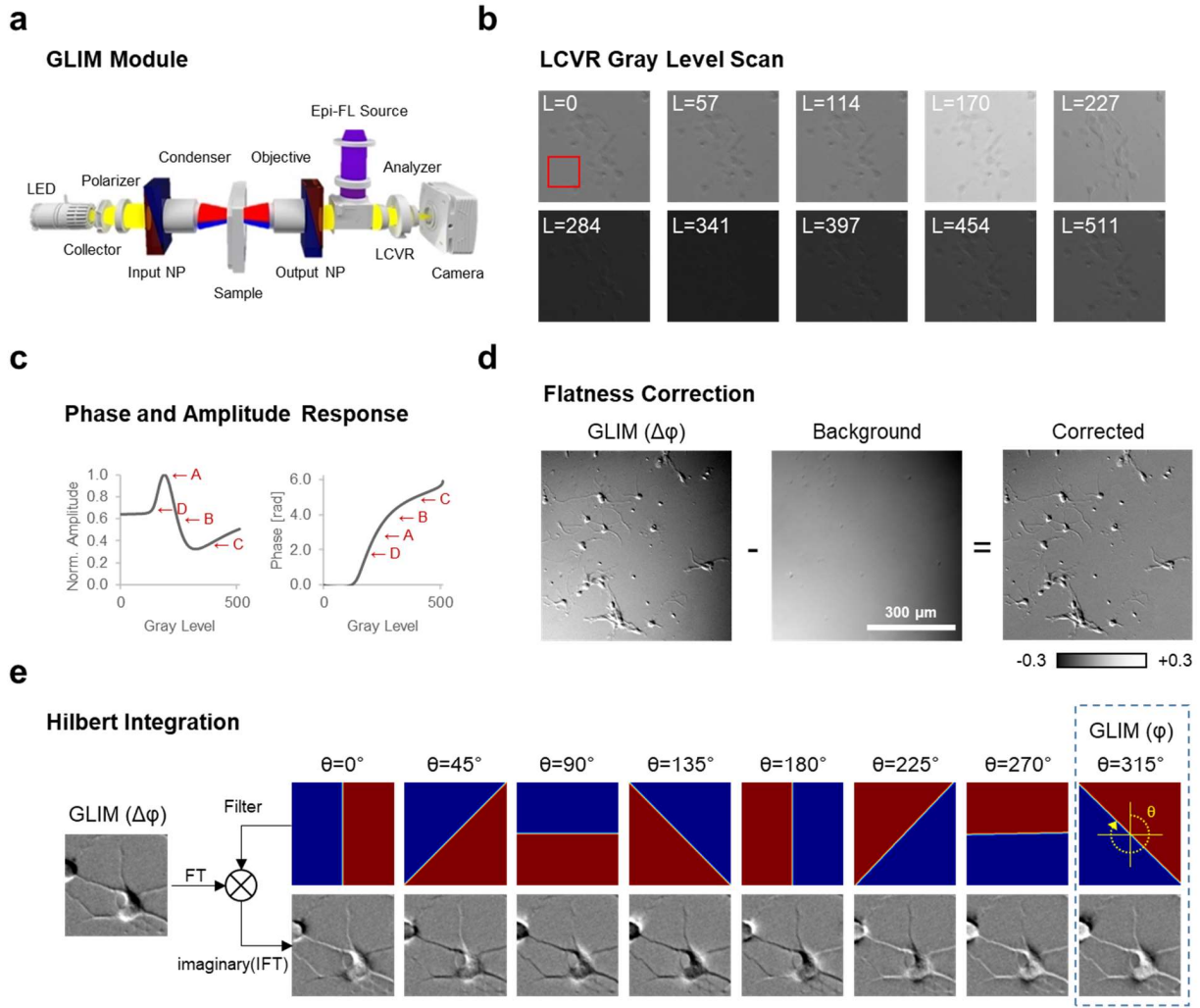

**Supplementary Figure 1. Design and calibration of the GLIM module. a,** Following the layout for DIC, the GLIM light path uses an input polarizer to introduce two beams that traverse the sample at laterally offset positions. These beams are combined in space by the output polarizer with the interference between the two polarizations controlled by a liquid crystal variable retarder mounted before the camera. **b,** To establish the relationship between the voltage on the modulator and the imparted phase shift, we perform a calibration procedure where "gray levels" with increasing phase shift are written to the modulator. **c,** Taking the average of these values over a simple-free region produces an amplitude modulation curve. Hilbert transform of which is the instantaneous phase associated with the modulator shift. **d**, To correct for background non-uniformities we perform background subtraction by removing an image corresponding to the average phase map acquired during the experiment. **e,** To integrate the phase we use a Hilbert transform where the Fourier transform of the GLIM image is multiplied by a filter corresponding to DIC shear direction. The imaginary part of the inverse transform contains the integrated phase. To determine the angle we evaluate all possible candidates (variations of ϕ shown) and select the one where all cellular structures introduce a positive phase shift (θ = 315° in this case).

Gradient light interference microscopy (GLIM) is a quantitative phase imaging technique that is constructed as an add-on module to a conventional DIC microscope[1]. In this work, we use a liquid crystal variable retarder based design first presented in [2]. GLIM is particularly well suited to reproduce turbid structures but also benefits from a relatively unobstructed fluorescent light path.

As shown in Supporting Fig. 1a, the GLIM light path begins with an incoherent light source. In this work, we use a red LED centered at 623 nm at the lowest power setting (Thorlabs, SOLIS-623C). The red color helps avoid harmful UV radiation and the ~50 nm bandwidth spectrum improves modulator contrast compared to a white light source. For imaging, we used a DIC microscope, which splits light into two beams that traverse the sample at slightly offset positions. In our implementation, we used an Axio Observer Z1 (Zeiss) equipped with an automated stage and mini-incubator. All imaging was performed on a 20x/0.8 objective with a fully open condenser (DICII/0.55). To obtain the quantitative phase information, we remove the output polarizer (in the filter cube) and place a liquid crystal variable retarder (LCVR, LVR-100-IR, Meadowlark) and polarizer immediately before the camera.

The LCVR is used to introduce a controlled phase shift between interfering beams in DIC which can then be used to recover the phase information. To establish the relationship between the gray-level and the phase shift we perform a calibration procedure where we acquire a sequence of 512 increasing voltages, conventionally referred to as "gray levels". In general, the calibration should be run for each light source, and this sequence resembles what would occur when rotating the de Sénarmont prism (Supporting Fig. 1b). To determine the relationship between gray-level and effective phase shift, we use a sample-free portion of the image and project the spatial average value as a function of gray-level (Supporting Fig. 1c). The instantaneous phase is recovered by the Hilbert transform of the normalized amplitude signal which results in the imaginary portion of the complex analytic signal, the argument of which is the phase. To obtain the final reconstruction formula we take the largest peak ("A") and find phase shifts that are -90°, 180°, 270° from that point. To account for discretization errors we use a phase reconstruction formula that recovers the phase when points that are slightly off from 90° intervals[2].

As the GLIM image is a measurement of the phase between DIC beams, it resembles a derivative of the scattering potential rather than the actual potential. To recover the phase shift associated with the object we remove the artificial slant like background and divide by the previously measured DIC shear distance (0.3 μm). In this work, we obtain a background image by averaging together all images in the time-lapse sequence (Supporting Fig. 1d), and for real-time imaging when setting up the experiment, we used Fourier filtering[3].

Numerical integration is performed using a Hilbert transform, which is good at preserving high-frequency information while avoiding the need to measure an impulse response or use a regularizer. Comparing the Hilbert transform to more formal integration methods, we note that the principle distortion is in low-frequencies which lay outside GLIM's frequency response[4]. To determine the angle of integration precisely we use an exhaustive approach where we generate 360 images corresponding to different integration angles (Supporting Fig. 1e). The correct angle of integrations can be visually identified by choosing an image where all cellular structures appear as positive phase shifts (315 degrees in this case). In principle, it should be possible to read the angle from the frequency spectra or simply determine the shear angle from the instrument.

**Supplementary Note 2: Cell-culture and immunostaining**

All procedures involving animals were reviewed and approved by the Institutional Animal Care and Use Committee at the University of Illinois Urbana-Champaign and conducted per the guidelines of the U.S National Institute of Health (NIH). Primary dissociated hippocampal neurons were prepared from the hippocampi which were dissected from Sprague-Dawley rat embryos at embryonic day 18 as described[5] and plated on to a glass-bottom multiwell plate (Cellvis, P06-20-1.5-N) that was functionalized with poly-D-lysine (0.1 mg/ml; Sigma-Aldrich). Hippocampal neurons were initially incubated with a plating medium containing 86.55% MEM Eagle's with Earle's BSS (Lonza), 10% Fetal Bovine Serum (re-filtered, heat-inactivated; ThermoFisher), 0.45% of 20% (wt./vol.) glucose, 1x 100 mM sodium pyruvate (100x; Sigma-Aldrich), 1x 200 mM glutamine (100x; Sigma-Aldrich), and 1x Penicillin/ Streptomycin (100x; Sigma-Aldrich) to be attached to the surface of the well plates. Neurons were plated at 300 cells/mm$^2$, although the achieved density was calculated on a per-tile basis using image analysis. After three hours of incubation (37°C and 5% $CO_2$), the plating media was aspirated and replaced with maintenance media containing Neurobasal™ growth medium (ThermoFisher) supplemented with B-27 (ThermoFisher), 1% 200 mM glutamine (ThermoFisher) and 1% penicillin/streptomycin (ThermoFisher) at 37 °C, in the presence of 5% $CO_2$. Hippocampal neurons were maintained for three days before imaging.

Timelapse microscopy was performed using the plate reader instrumentation developed in[4]. Imaging was performed on three wells with a 10 x 10 mosaic performed every 50 minutes. To avoid altering the polarization in DIC we used a glass top for imaging (Cellvis, L001). At the end of the experiment, cells were stained and imaged using co-localized fluorescence and GLIM imaging.

In this study, antibodies for Tau (Abcam, ab80579) and MAP2 (Abcam ab32454) were used for immunostaining of axons and dendrites, respectively, following the established protocol from the ThermoFisher and Abcam[6]. In brief, neurons were fixed with freshly prepared 4% paraformaldehyde (PFA) for 15 minutes following 0.5% Triton-X for 10 minutes and 2% BSA for 30 minutes incubation. Hippocampal neurons were incubated for 8 hours at 4°C with anti-Tau antibodies that was diluted to 1:250 in 1% bovine serum albumin (BSA, ThermoFisher). After washing, neurons were exposed for 8 hours at 4°C to goat anti-mouse secondary antibody (Abcam, ab205719) which was diluted to 1:500 in 1% BSA. Hippocampal neurons were then incubated in anti-MAP2 antibody (1:500 dilution) in 1% BSA for 8 hours, followed by goat anti-rabbit secondary antibody (Abcam, ab205718, 1:1000 dilution) in 1% BSA for 8 hours at 4°C.

We found that the fluorescent signal covered roughly 7% of the histogram (~ 150 counts from 2047 available). To improve this dynamic range, and we used a combination of 2 x 2 binning and 10 frames averaging for each fluorescence channel (Photometrics, BSI Prime). This fluorescence acquisition is roughly 70x slower than the time to acquire the GLIM image.

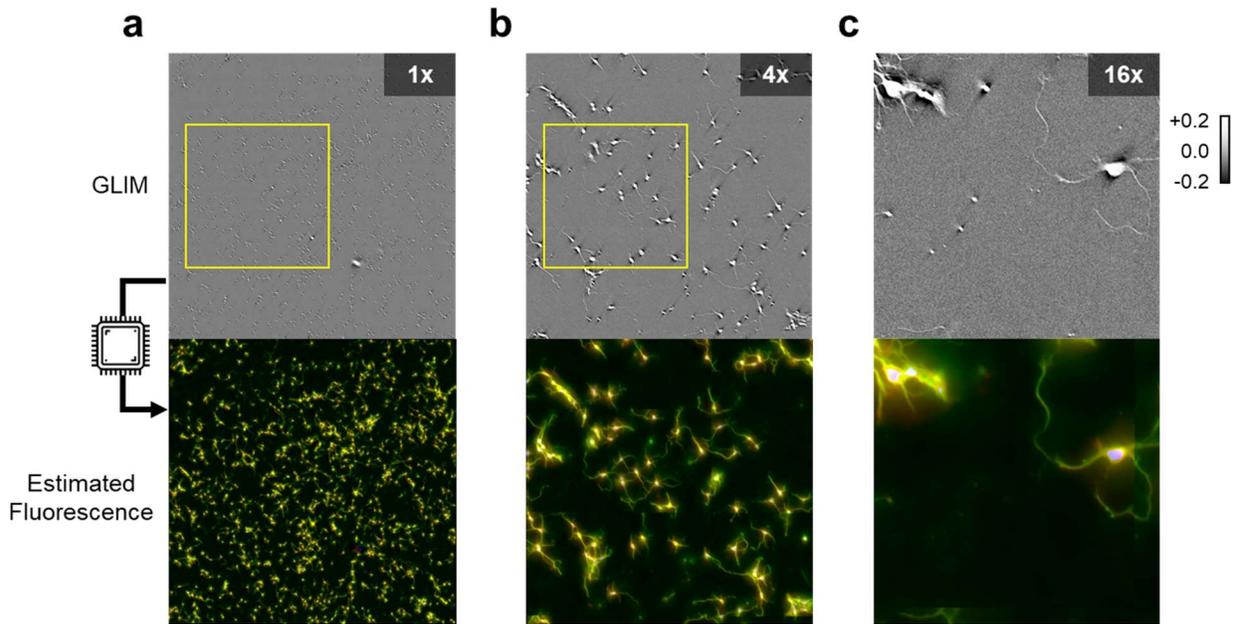

**Supplementary Figure 2. Cellular level detail in GLIM and PICS. a-c** Representative field of view shown alongside the estimated fluorescent signal at increasing zoom levels (20x/0.8).

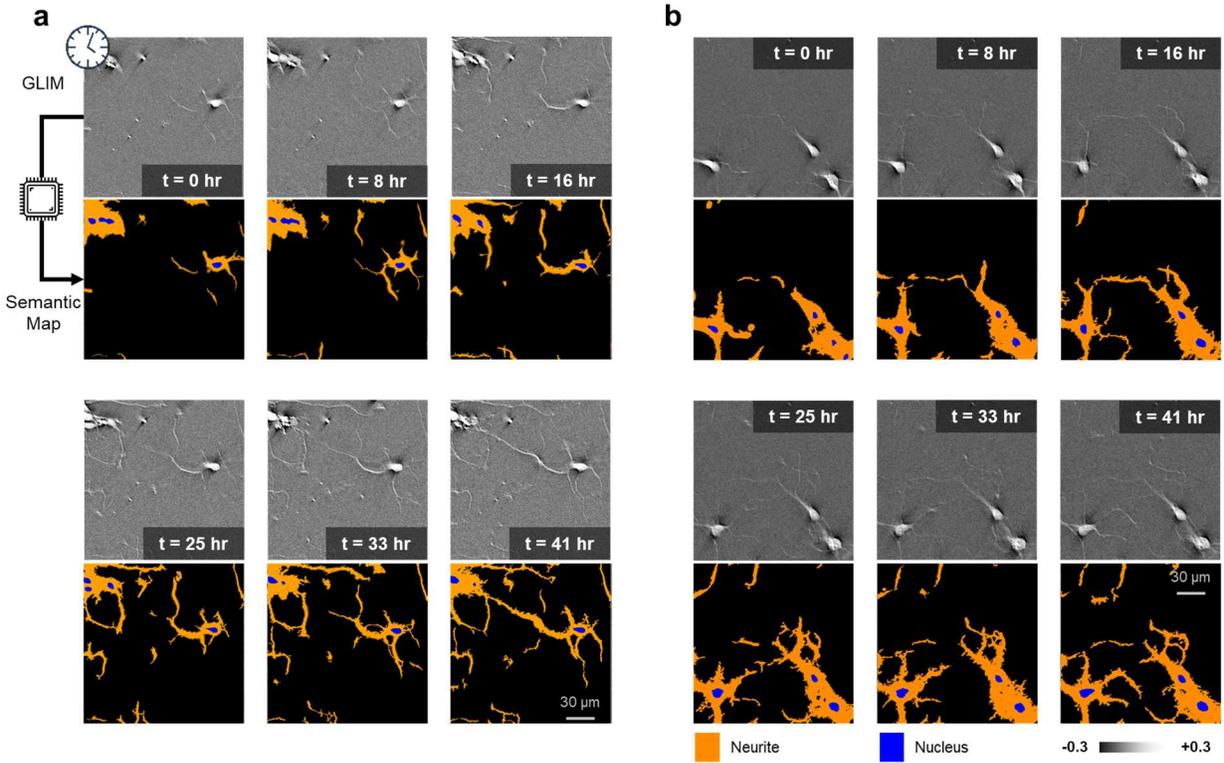

**Supplementary Figure 3. Semantic segmentation and time-lapse development of the neuronal arbor.** Two representative fields of view along with associated semantic segmentation map (20x/0.8).

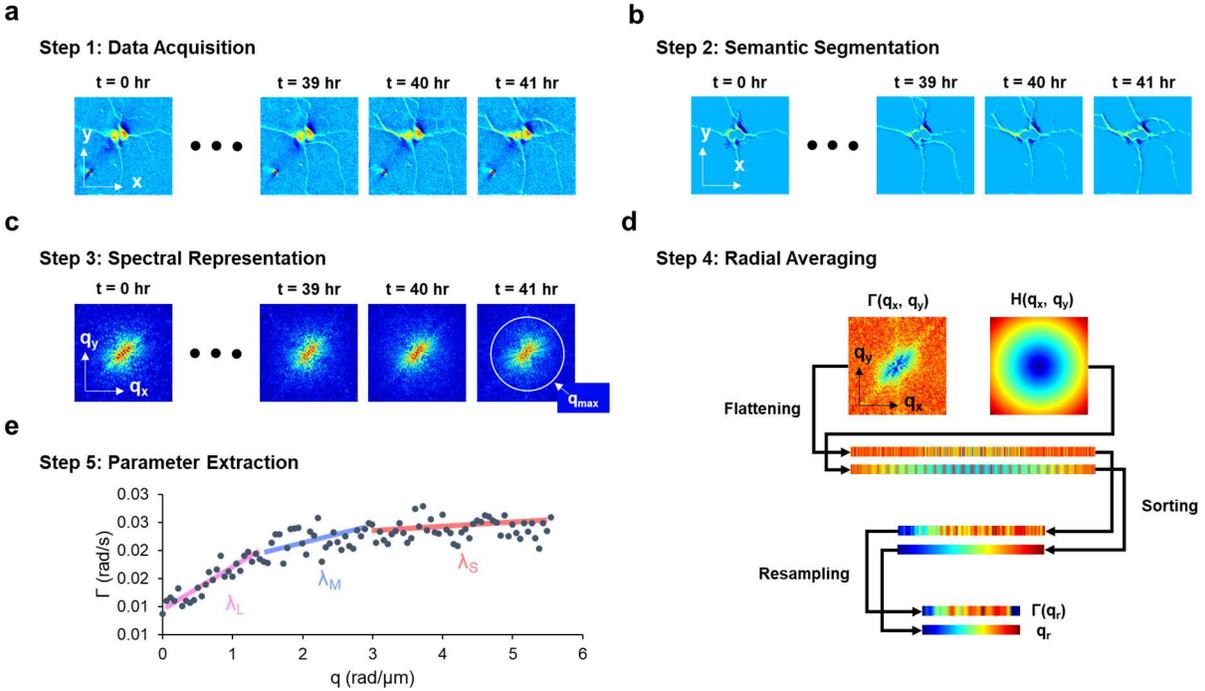

**Supplementary Figure 4. Processing for dispersion-relation phase spectroscopy (DPS). a,** The DPS procedure estimates the spread of the advection coefficients from a series of phase (shown) or fluorescent measurements. **b,** To apply the semantic segmentation to the phase image, we zero all values outside the labeled regions. **c,** Next, we take the Fourier transform of the sequence on a per tile basis. As expected, the resolution limit ($q_{max}$) appears as a disk in the frequency domain. **d,** Next we estimate the variance of the autocorrelation using a finite difference scheme to obtain $\Gamma(q_x,q_y)$. As cellular transport is assumed to be anisotropic we perform a radial average by indexing $\Gamma$ with q [the modulus of $\mathbf{q}=(q_x,q_y)$] followed by a resampling step to reduce the radially sorted data from ~1 megapixels to a more manageable 1,000 samples. **f,** Finally, we perform a linear curve fit over the ranges corresponding to "small" ($\lambda_S$), "medium" ($\lambda_M$), and "large" ($\lambda_L$) scale features.

**Video 1**

GLIM and PICS estimated fluorescence for Tau/MAP2/DAPI for the first hippocampal culture. Imaging was performed for 41 hours at (20x/0.8).

**Video 2**

GLIM and PICS estimated fluorescence for Tau/MAP2/DAPI for the second hippocampal culture. Imaging was performed for 41 hours at (20x/0.8).

**Video 3**

GLIM and PICS estimated fluorescence for Tau/MAP2/DAPI for the third hippocampal culture. Imaging was performed for 41 hours at (20x/0.8).